\documentclass[a4paper,10pt]{article}
\usepackage{amssymb,amsmath,libertine}
\usepackage[left=2.5cm,right=2.5cm,bottom=2.5cm,top=2.5cm]{geometry}

\usepackage[utf8]{inputenc}
\usepackage[english]{babel}
\usepackage[T1]{fontenc}

\usepackage{bm}
\usepackage{color}
\usepackage{authblk}

\usepackage{hyperref}

\newcommand{\de}{\partial}

\newcommand{\vc}[1]{\mathbf{#1}}
\newcommand{\vt}[1]{\mathsf{#1}}
\newcommand{\tsp}{\mathsf{T}}

\newcommand{\dvg}{{\mathrm{div}}}

\begin{document}

\title{Rejection of the principle of material frame indifference}

\author{Giulio G.~Giusteri}
\affil{Mathematical Soft Matter Unit, Okinawa Institute of Science and Technology Graduate University\\ 1919-1 Tancha, Onna, Okinawa, 904-0495, Japan\\
\textrm{giulio.giusteri@oist.jp}}
\date{February 4, 2017}


\maketitle

\begin{abstract}
The principle of material frame indifference is shown to be incompatible with the basic balance laws of continuum mechanics. In its role of providing constraints on possible constitutive prescriptions it must be replaced by the classical principle of Galilean invariance.
\end{abstract}

\section*{}

The principle of material frame indifference as stated by Truesdell \& Noll~\cite{TruNol65} is considered an important principle in continuum mechanics. It has been chiefly used to identify constraints on possible constitutive prescriptions. Here, via an elementary argument, it is shown to be incompatible with the basic balance laws of continuum mechanics.
The argument also indicates that, in its role of guidance for establishing physically acceptable constitutive prescriptions, it must be replaced by the classical Galilean invariance. In fact, all of the physically correct consequences of frame indifference that are presented in countless research papers and books can be obtained by only requiring Galilean invariance, as can be checked by inspecting the corresponding proofs. On the other hand, those consequences of material frame indifference that do not descend from Galilean invariance do not constitute physically relevant statements.

No matter how the principle of material frame indifference is worded or described, its mathematical formulation says that the response functions relating kinematic descriptors of a continuum with energy, stress, or other dynamic quantities must obey the transformation rules
\begin{equation}\label{eq:tr}
f^*(\vc x^*)=f(\vc x)\,,\qquad
\vc v^*(\vc x^*)=\vt Q\vc v(\vc x)\,,\qquad\mathrm{and}\qquad
\vt T^*(\vc x^*)=\vt Q\vt T(\vc x)\vt Q^\tsp\,,
\end{equation}
stated for scalar, vector, and tensor fields, respectively. 
The asterisk indicates a change of observer, defined as a transformation in the Euclidean ambient space such that a position vector $\vc x$ transforms according to
\begin{equation}\label{eq:obs}
\vc x^*=\vc c(t)+\vt Q(t)\vc x\,,
\end{equation}
where $\vc c$ is any time-dependent vector, and $\vt Q$ is any time-dependent orthogonal transformation.
Moreover, it is assumed that scalar fields always obey the foregoing transformation rules.

Let us now consider the well-known theory of isentropic fluids. The fields that describe the state of the system are the mass density $\rho$ and the fluid velocity $\vc u$. The basic balance laws of continuum mechanics for such systems in the absence of external body forces or mass sources read
\begin{equation}
\frac{\de\rho}{\de t}+\dvg\rho\vc u=0\,,
\end{equation}
and
\begin{equation}
\rho\vc a=\rho\bigg(\frac{\de\vc u}{\de t}+(\vc u\cdot\nabla)\vc u\bigg)=\dvg\vt T\,,
\end{equation}
where the constitutive prescription for the Cauchy stress tensor is
\begin{equation}\label{eq:T}
\vt T=-p\vt I\,.
\end{equation}
In this relation, the stress is proportional to the identity tensor $\vt I$ via a pressure field $p$ determined by the values of $\rho$ through the constitutive prescription
\begin{equation}\label{eq:p}
p(\vc x)=f(\rho(\vc x))\,,
\end{equation}
where $f$ is an invertible real function of a real variable.
Since the scalar function $f$ does not depend explicitly on $\vc x$, it is immediate to verify that the constitutive prescriptions \eqref{eq:T}--\eqref{eq:p} comply with the principle of material frame indifference.

If we consider a solution $(\rho,\vc u)$ of the evolution equations, together with \eqref{eq:T}--\eqref{eq:p}, the balance of momentum takes the form
\begin{equation}\label{eq:1}
\rho\vc a+\nabla f(\rho)=\vc 0\,.
\end{equation}
We now apply a change of observer such that $\vc c$ vanishes identically. Denoting time derivatives with a superimposed dot, introducing the antisymmetric tensor $\vt\Omega(t)=\dot{\vt Q}(t)\vt Q^\tsp(t)$, and recalling that $\nabla^*=\vt Q\nabla$ and that scalar fields obey \eqref{eq:tr}$_1$, we have
\begin{equation}
(\rho\vc a)^*=\rho\vc a^*=\rho[\vt Q\vc a+2\dot{\vt Q}\vc u+(\dot{\vt\Omega}-\vt\Omega^2)\vc x^*]
\end{equation}
and
\begin{equation}
[\nabla f(\rho)]^*=\vt Q\nabla f(\rho)\,.
\end{equation}
Since the null vector field $\vc 0$ is the gradient of any constant scalar field, we have $\vc 0^*=\vt Q\vc 0=\vc 0$ and the foregoing relations yield
\begin{equation}\label{eq:2}
\rho[\vt Q\vc a+2\dot{\vt Q}\vc u+(\dot{\vt\Omega}-\vt\Omega^2)\vc x^*]+\vt Q\nabla f(\rho)=\vc 0\,.
\end{equation}

Multiplying now \eqref{eq:1} by $\vt Q$ and subtracting it from \eqref{eq:2} we obtain
\begin{equation}\label{eq:3}
2\dot{\vt Q}\vc u+(\dot{\vt\Omega}-\vt\Omega^2)\vc x^*=\vc 0\,,
\end{equation}
which holds true for any $\vc x^*$, any solution of the evolution equations, and any choice of the time-dependent orthogonal transformation $\vt Q$. The simple choice of hydrostatic solutions, for which $\vc u=\vc 0$, and of a constant nonvanishing spin tensor $\vt\Omega$ leads to
\begin{equation} 
\vt\Omega=\vt 0\,,
\end{equation}
which clearly contradicts our choice of a nonvanishing $\vt\Omega$. We thus conclude that the principle of material frame indifference is inconsistent with the basic balance laws of continuum mechanics. In other words, the transformation of the balance equations under general changes of observer of the form \eqref{eq:obs} is not compatible with the invariance of scalar fields assumed by the principle of material frame indifference.

On the contrary, no contradiction arises if we restrict changes of observer to the case of a time-independent $\vt Q$, since \eqref{eq:3} would be trivially satisfied in those cases. This, together with an analogous argument in which the change of observer is defined with a vector $\vc c$ proportional to $t^2$ (uniform acceleration), shows that we must require only Galilean invariance of the constitutive prescriptions to avoid such contradictions. 

In fact, the general argument behind the foregoing result is that, since the field $\rho\vc a$ transforms according to \eqref{eq:tr} only if we confine ourselves to Galilean changes of observers, we cannot require a stronger invariance to the fields that are equated to $\rho\vc a$ in the fundamental balance equations. This shows that the principle of material frame indifference is always in contradiction with the basic balances.
It should be noted that, consistent with this conclusion, M\"uller \cite{Mul72} already showed that the principle of material frame indifference cannot always be valid, since it is not compatible with the derivation from kinetic theory of the continuum equations for certain gases.
Based also on the general validity and importance of Galilean invariance in nonrelativistic mechanics, it is clear that it should replace material frame indifference in constraining the selection of constitutive prescriptions.

\subsection*{Acknowledgement}
The author gratefully acknowledges support from the Okinawa Institute of Science and Technology Graduate University with subsidy funding from the Cabinet Office, Government of Japan. The author also thanks I.~Murdoch and B.~Svendsen for providing useful comments on the manuscript.


\end{document}